# Modeling Growth and Plasma Oxygen Effects on Metal Purity in Platinum EBID

*Marianna D'Amato[1], Linda Piscopo[2], Hanna Le Jeannic[1,‡], Alberto Bramati[1,‡], Antonio Balena[1,2,‡,*]*

[1.] *Laboratoire Kastler Brossel, Sorbonne Université, CNRS, ENS-PSL Research University, Collège de France, Paris, France*

[2.] *Istituto Italiano di Tecnologia, Center for Biomolecular Nanotechnologies, Arnesano (LE), Italy*

[‡] *Co-last authors*

[*] *Corresponding author: antonio.balena@iit.it*

## Abstract

Electron Beam-Induced Deposition (EBID) enables site-specific nanofabrication but suffers from significant carbon contamination, limiting its applicability in plasmonics, nanoelectronics, and sensing. In this study, we investigate the relationship between EBID process parameters such as beam current, acceleration voltage, and dwell time, and the platinum-to-carbon composition of deposited nanostructures. Using Energy Dispersive X-Ray Spectroscopy (EDX), we establish a hindered exponential growth model that correlates deposit composition with fabrication conditions. To enhance metal purity, we apply plasma oxygen treatment, exposing EBID deposits to a 30 W plasma for 30 minutes in a tabletop plasma generator. Post-treatment EDX analysis confirms a systematic increase in platinum content, while SEM inspection reveals nanostructure shrinkage due to carbon removal. This work aims to provide a framework for optimizing EBID fabrication and post-processing strategies to enhance material performance.

## 1. Introduction

Electron Beam Induced Deposition (EBID) is a maskless, direct-write nanofabrication technique that relies on the electron-beam-induced dissociation of a volatile organometallic precursor injected inside the chamber of a scanning electron microscope (SEM)[1,2]. When the electron beam interacts with the substrate, it generates secondary electrons that decompose the precursor molecules, leading to the formation of a solid deposit, while volatile by-products are removed via the vacuum system. Thanks to its high spatial resolution, which allows for achieving direct-write nanostructure formations down to a few nanometers[3,4], EBID found applications in plasmonics[5], nanoelectronics[6], and nanostructured sensors[7,8]. In a recent work,[9] we introduced Blurred EBID (BEBID), a modified approach that enhances EBID's versatility targeting delicate substrates. By defocusing the electron beam, we showed that BEBID enabled high-precision deposition of platinum (Pt) nanostructures on tapered optical nanofibers while minimizing mechanical stress and preserving substrate integrity.

However, EBID suffers from a major limitation: significant carbon (C) contamination in the deposited material. This contamination arises due to incomplete dissociation of the precursor or the trapping of the volatile by-product within the growing deposit[10,11]. The residual C content can severely impact the optical, electrical, and mechanical properties of a deposited nanostructure, leading to reduced electrical conductivity, increased resistivity, and altered optical response[12]. Over the years, numerous studies have sought to characterize the properties of materials deposited by EBID and to understand the influence of fabrication parameters and of purification strategies on their composition. One of the earliest studies on EBID-deposited Pt structures using the same organometallic precursor as in this work was conducted by H. W. P. Koops and colleagues[13]. This study already provided insights into how electron beam current influences the relative Pt concentration in the deposited structure. Later, Belić and co-authors[14] expanded on these findings by developing a more detailed description of the influence of fabrication parameters on metal concentration.

To mitigate the C contamination limitation, and to improve the metal purity of EBID structures, several post-deposition purification strategies have been proposed. Thermal annealing[15,16] can effectively remove C contamination but can induce structural deformation and diffusion mechanisms in the deposit. Post-deposition electron irradiation[17,18] enables in-situ purification but requires high-energy irradiation, which may alter the morphology of the nanostructures. Gas-assisted EBID (GA-EBID)[19] improves purity by introducing co-reactive gases during growth but requires complex modifications of the gas delivery setup. Alternatively, Plasma oxygen treatment offers a practical, non-destructive and scalable approach to improving EBID deposit purity while preserving nanoscale precision. During plasma exposure, reactive oxygen species facilitate the oxidation of C-rich contaminants, which are subsequently removed in the form of volatile CO and $CO_2$ molecules[11], thus enhancing metal purity while preserving nanostructural fidelity[20]. For example, Belić and co-authors[14] also investigated the effects of post-exposure to oxygen plasma on metal concentration in EBID structures using an organometallic gold precursor.

Despite these seminal works, no comprehensive study has systematically examined the interdependence of Pt concentration on multiple fabrication parameters. This work aims to address this gap by providing a detailed analysis of how electron beam conditions influence material composition in EBID-deposited Pt nanostructures, and to evaluate the effectiveness of plasma oxygen treatment in enhancing purity while maintaining structural integrity. First, we analyze how electron beam parameters (current, accelerating voltage, and dwell time) affect the composition of the deposits, specifically the platinum-to-carbon (Pt-to-C) fill factor, using Energy Dispersive X-Ray Spectroscopy (EDX) measurements. Second, we evaluate the effectiveness of plasma oxygen treatment by exposing EBID deposits to a 30 W plasma for 30 minutes in a tabletop plasma generator, a widely available system in many research and fabrication facilities. The purification effect was assessed through EDX measurements, quantifying changes in the Pt-to-C ratio. Additionally, SEM inspection of nanostructures revealed a systematic reduction in feature dimensions due to C removal. This study aims to provide a comprehensive framework for optimizing fabrication parameters and purification strategies. The resulting composition, derived from combined geometric and compositional analysis, may further serve as a foundation for optimizing optical and electronic properties via effective medium approximations[21].

## 2. Results

### 2.1. Influence of EBID Parameters on Pt Composition

**Figure 1A** shows a SEM micrograph of the sample designed for EBID characterization. The sample consists of three 4 × 4 arrays of squares patterned on a ~100 nm gold layer using Focused Ion Beam Milling (FIBM). At the center of each square, a 1.6 × 1.6 × 0.5 µm³ EBID cube is grown. Each cube corresponds to a unique combination of accelerating voltage $V_{acc}$, beam current $I_{beam}$, and dwell time $\tau_{dwell}$. **Figure 1B** provides a higher-magnification SEM micrograph of a typical structure, while **Figure 1C** illustrates a cross-sectional schematic of the sample layout. Each of the three arrays is fabricated at a different acceleration voltage $V_{acc}$ of 1, 3, or 5 kV. As illustrated in **Figure 1A**, the beam current $I_{beam}$ (vertical axis) varies among 5.4, 23, 86, and 340 pA, while the dwell time $\tau_{dwell}$ (horizontal axis) ranges from 0.4 to 1.0 µs in 0.2 µs increments. For the lowest beam current (5.4 pA), only dwell times of 0.4 µs and 0.8 µs are used, because fabricating a single structure at this current requires over three hours, making it impractical for resource-efficient fabrication. In total, 42 unique parameter combinations are therefore investigated, providing a strong spanning range for evaluating the influence of beam conditions on the EBID structures' composition. For each structure, an EDX spectrum is acquired. A typical spectrum is shown in **Figure 1D**. revealing characteristic peaks of the substrate[22] (soda–lime glass, SLG), along with a peak attributed to gallium ion implantation[23] from the FIBM process. Pt and C peaks can also be identified, and are used to quantify the composition of the deposited structures.

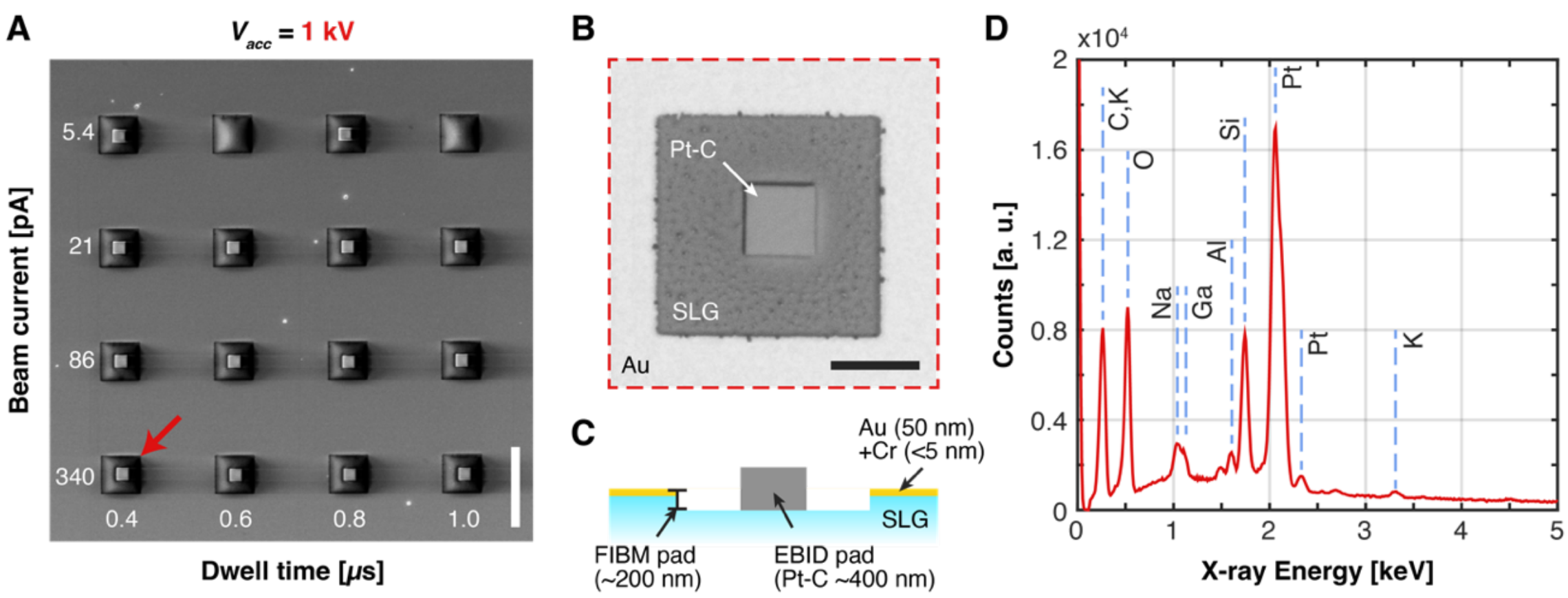

***Figure 1. Characterization of the sample used for EBID deposition and analysis. (A)*** *SEM micrograph of a representative 4 × 4 square array patterned via Focused Ion Beam Milling (FIBM) onto a 170-µm-thick SLG coverslip coated with ~100 nm of gold and a 5 nm chromium adhesion layer. Each square contains a centrally positioned 1.6 × 1.6 × 0.5 µm³ EBID-grown cube, fabricated using a unique combination of deposition parameters.* ***(B)*** *High-magnification SEM micrograph of a single EBID structure.* ***(C)*** *Schematic cross-sectional view of the sample design for characterization.* ***(D)*** *Representative EDX spectrum of an EBID-deposited structure, showing characteristic peaks of the SLG substrate, gallium ion implantation from FIBM, and the primary elements of interest: Pt and C.*

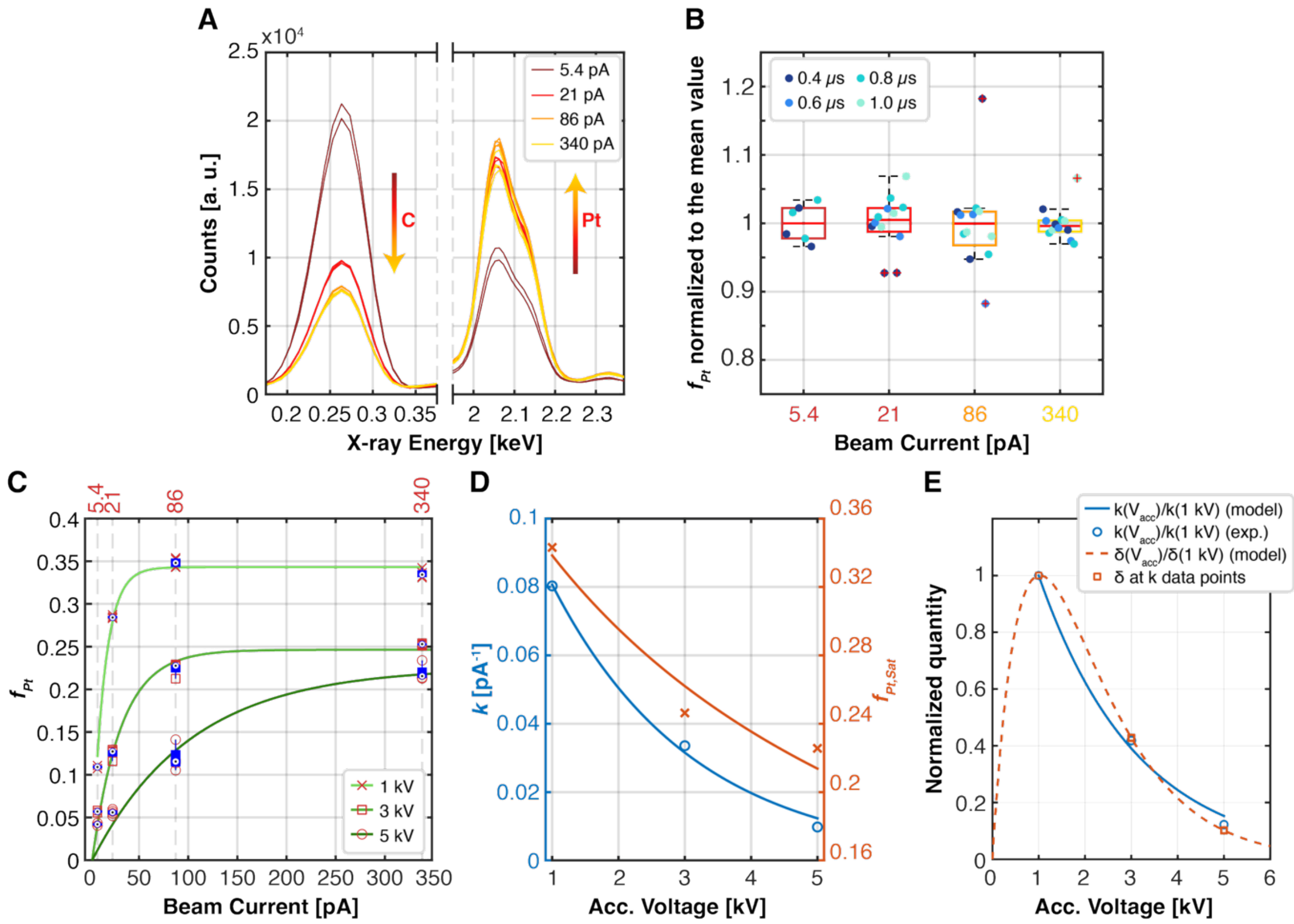

***Figure 2. Analysis of the dependence of the Pt fill factor $f_{Pt}$ on the electron beam parameters. (A)*** *Representative EDX spectra for structures fabricated at $V_{acc}$ = 1 kV, showing the evolution of Pt and C peak intensities as a function of beam current $I_{beam}$.* ***(B)*** *Boxplot representation of $f_{Pt}$ grouped by $I_{beam}$ and normalized to their mean value at a fixed $V_{acc}$, demonstrating the negligible influence of dwell time $\tau_{dwell}$ on composition.* ***(C)*** *Dependence of $f_{Pt}$ on $I_{beam}$ for different acceleration voltages $V_{acc}$, with data fitted to a hindered exponential growth model.* ***(D)*** *Voltage dependence of the saturation Pt fill factor $f_{Pt,Sat}$ and growth rate constant k. The least squares fits show distinct trends for both parameters, as discussed in the main text.*

The effect of the beam parameters on the relative intensities of C and Pt peaks in the EDX spectra is illustrated in **Figure 2A** for $V_{acc}$ = 1 kV, focusing on the $K\alpha$ x-ray line of C (0.277 keV) and on the $M\alpha$ x-ray line of Pt (2.048 keV), which is the most intense Pt-line in the spectra. Qualitatively, an increase in $I_{beam}$ results in a decrease in the C peak intensity, accompanied by a simultaneous increase in the Pt peak intensity, up to a saturation.

The EDX spectra are used to determine the relative Pt fill factor, $f_{Pt}$, defined as the volumetric ratio of Pt-to-C content in the deposits (see *Methods*). **Figure 2B** presents a boxplot illustrating the dependence of $f_{Pt}$ on $\tau_{dwell}$. The $f_{Pt}$ values are grouped by $I_{beam}$ and normalized by their mean value at a fixed $V_{acc}$. Each boxplot displays the median as the central mark, with the lower and upper edges representing the 25th and 75th percentiles, respectively. Whiskers extend to the most extreme data points that are not classified as outliers, while outliers are plotted individually using the '+' marker symbol. The data are symmetrically distributed around the median, with relative fluctuations in $f_{Pt}$ remaining below ±5%. This suggests that, within the explored range, $\tau_{dwell}$ has no systematic influence on $f_{Pt}$. Consequently, in the following analysis, data are grouped by $I_{beam}$, neglecting variations due to $\tau_{dwell}$.

The dependence of $f_{Pt}$ on $I_{beam}$, is reported in **Figure 2C** with data grouped by $V_{acc}$. Similar to the trend observed for $V_{acc}$ = 1 kV, $f_{Pt}$ increases with $I_{beam}$ for all the $V_{acc}$ values. For each $V_{acc}$, the data can be well described using a hindered exponential growth model, expressed as:

$$f_{Pt}(I_{beam}) = f_{Pt,Sat} \cdot \left(1 - e^{-kI_{beam}}\right), \tag{1}$$

where $f_{Pt}(I_{beam})$ is the measured Pt fill factor at a given $I_{beam}$, $f_{Pt,Sat}$ represents the saturation Pt fill factor, and $k$ [$A^{-1}$] is a growth rate constant that sets how rapidly $f_{Pt}$ approaches $f_{Pt,Sat}$ as $I_{beam}$ increases. The extracted parameters are listed in **Table 1**.

| $V_{acc}$ [kV] | $f_{Pt,Sat}$ | $k$ [$pA^{-1}$] |
|---|---|---|
| 1 | 0.3431 (0.3363, 0.3498) | 0.0803 (0.0734, 0.0874) |
| 3 | 0.2463 (0.2369, 0.2557) | 0.0336 (0.0288, 0.0383) |
| 5 | 0.2256 (0.1994, 0.2516) | 0.0098 (0.0067, 0.0129) |

***Table 1. Hindered exponential growth fitting model.*** *Extracted values of $f_{Pt,Sat}$ and $k$ for each value of $V_{acc}$. In parentheses the 95% confidence bounds are reported.*

Both $f_{Pt,Sat}$ and $k$ exhibit a marked dependence on $V_{acc}$. The explicit dependence of $f_{Pt,Sat}$ and $k$ on $V_{acc}$ is summarized in **Figure 2D**. A least squares fit reveals that $f_{Pt,Sat}$ is well described by a simple rational function which captures the rapid decrease of the maximum achievable Pt content as $V_{acc}$ increases:

$$f_{Pt,Sat}(V_{acc}) = \frac{A}{(1 - BV_{acc})^{n}}, \tag{2}$$

where $A$ = 0.3962 could be interpreted as the saturation Pt fraction when $V_{acc}$ approaches zero voltage, $B$ = 0.17 $kV^{-1}$ quantifies the rate at which $f_{Pt,Sat}$ decreases with increasing $V_{acc}$, and $n$ = 1 suggests a direct inverse dependence on (1 - $BV_{acc}$). **Equation 2** indicates that $f_{Pt,Sat}$ decreases non-linearly with voltage, initially dropping rapidly before approaching an asymptotic limit as $V_{acc}$ increases.
In parallel, the growth rate constant $k$ follows an exponential decay with $V_{acc}$, expressed as:

$$k(V_{acc}) = k_0 e^{-\alpha V_{acc}}, \tag{3}$$

where $k_0$ = 0.129 $pA^{-1}$ is the characteristic growth rate as $V_{acc}$ approaches zero, and $\alpha$ = 0.47 $kV^{-1}$ is a decay constant that determines how sensitively $k$ decreases with increasing $V_{acc}$. This relationship suggests that the approach to saturation becomes progressively slower at higher voltages.
Since precursor conditions remain constant across all the fabrications, both trends highlight the central role of the near-surface secondary-electron (SE) population in driving precursor dissociation in EBID. Indeed, at low $V_{acc}$, the interaction volume is shallow, and the SE yield is maximized near the surface, favoring more effective precursor decomposition and reducing the retention of C-based ligands, thus a larger $f_{Pt,Sat}$. Additionally, reduced forward scattering at lower voltages ensures more energy is deposited in the precursor-rich region, further enhancing metal incorporation. Conversely, at high $V_{acc}$, deeper electron penetration reduces SE availability at the surface, limiting precursor dissociation efficiency and lowering $f_{Pt,Sat}$. These findings align with previous studies on electron interaction depth and SE-induced deposition efficiency in EBID, such as those by Koops et al. (1994)[24], which demonstrated that higher SE generation at lower voltages enhances metal purity in EBID deposits, and Van Dorp & Hagen (2008)[25], which reviewed the dominant role of SEs in precursor dissociation efficiency. Furthermore, modeling by Silvis-Cividjian et al. (2005)[26] showed that electron penetration depth at higher voltages can lead to precursor transport limitations, further reinforcing the observed reduction in $f_{Pt,Sat}$ at high $V_{acc}$.

By construction, $k$ can be interpreted as an effective kinetic parameter that aggregates how efficiently a given set of beam conditions converts incident primary electrons, through secondary electrons, into the reactive events that induce precursor dissociation and Pt incorporation into the growing deposit. To test this interpretation, we directly compare the extracted $k(V_{acc})$ values with a simple model for the SE yield $\delta(V_{acc})$. Our goal is not to establish an exact quantitative mapping between $k$ and $\delta$ but rather to test whether the monotonic decrease of $k$ with increasing accelerating voltage is at least semi-quantitatively compatible with the well-established decrease of $\delta$ in the 1-5 kV range. To this end, we adopt a minimal analytical form that reproduces the generic rise and fall of $\delta(E)$ with beam energy, characterized by a maximum at an effective energy $E_{max}$:[27,28]

$$\delta(E) \propto \left(\frac{E}{E_{max}}\right) e^{\left(1-\frac{E}{E_{max}}\right)}, \tag{4}$$

We then choose $E_{max}$ such that the normalized SE-yield curve $\delta(V_{acc})/\delta(1\text{ kV})$ best matches the normalized growth constant $k(V_{acc})/k(1\text{ kV})$ for $V_{acc}$ = 1, 3, and 5 kV. The best agreement is obtained for $E_{max} \approx 1$ keV, which is fully consistent with typical SE-yield maxima reported for metals and metal-containing surfaces[27]. The resulting modeled $\delta(V_{acc})$ reproduces both the direction and the relative magnitude of the decrease observed experimentally in $k(V_{acc})$ (**Figure 2E**).

This semi-quantitative agreement supports the view that the $V_{acc}$ dependence of $k$ is primarily governed by the underlying beam-matter interaction physics that controls the availability of reactive secondary electrons at the surface of the growing Pt–C nanostructures[29], while still acknowledging that $k$ also folds in additional contributions from precursor transport and surface kinetics.

These findings have direct implications for optimizing EBID fabrication parameters to achieve higher metal purity and deposition efficiency. Since lower $V_{acc}$ enhances precursor dissociation and increases metal incorporation, reducing $V_{acc}$ to around 1 kV may be beneficial when high-purity metallic deposits are required. However, operating at low $V_{acc}$ may also lead to increased electron scattering and broader feature sizes[25], which could be a limitation for ultra-high-resolution patterning. Conversely, higher $V_{acc}$ provides better beam focus and feature resolution but reduces SE-driven precursor dissociation, leading to lower $f_{Pt,Sat}$ and a slower approach to saturation.

### 2.2. Compositional effects of Plasma Oxygen Treatment

Following the characterization of as-deposited EBID structures, the effects of post-deposition oxygen plasma exposure were investigated to evaluate its role in enhancing metal purity and modifying material composition. The same sample was exposed to an oxygen plasma for 30 minutes at a power of 30 W. After treatment, the Pt fill factor $f_{Pt}$ was re-evaluated using EDX and compared to the pre-exposure values.

**Figure 3A** presents a representative EDX spectrum before and after plasma exposure for the structure fabricated at $V_{acc}$ = 1 kV, $I_{beam}$ = 340 pA, and $\tau_{dwell}$ = 1.0 µs. As expected from previous studies, it can be qualitatively observed that the relative intensity of the C *Kα* peak decreases with respect to the Pt *Mα* peak, indicating a relative increase in the metal proportion after plasma treatment. This effect is attributed to the oxygen plasma interacting with residual carbonaceous components in the EBID structure. The limited influence of dwell time on $f_{Pt}$ persists even after plasma treatment, as demonstrated in the boxplot in **Figure 3B**. Although the distribution of $f_{Pt}$ values around the median appears slightly broader compared to the pre-treatment state, the variation remains within 10%, with no clear functional dependence on $\tau_{dwell}$. **Figure 3C** presents the post-treatment $f_{Pt}$ values, which are fitted using the same hindered exponential growth model as in the pre-treatment analysis (**Equation 1**). The fitting results are reported in **Table 2**. A substantial increase in $f_{Pt}$ is observed at $V_{acc}$ = 1 kV, with $f_{Pt,Sat}$ that rises from 0.34 to 0.50, corresponding to a relative increase of nearly 50%. At $V_{acc}$ = 3 kV, $f_{Pt,Sat}$ also exhibits a notable increase of approximately 25%. In contrast, at $V_{acc}$ = 5 kV, the increase in $f_{Pt}$ is minimal, with $f_{Pt,Sat}$ remaining nearly unchanged relative to its pre-treatment value.

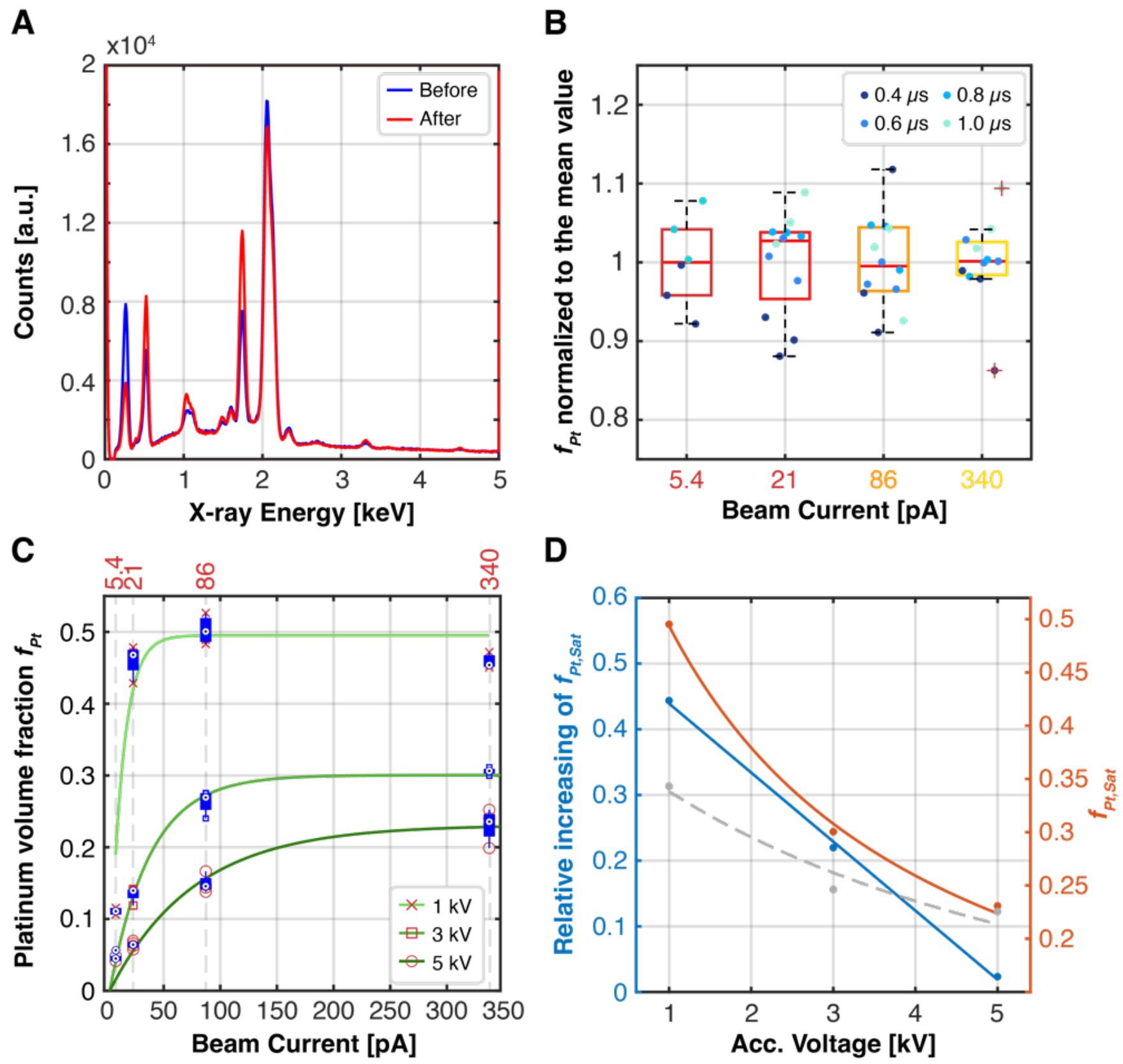

***Figure 3. Effects of oxygen plasma treatment on EBID-deposited structures. (A)*** *Representative EDX spectrum of a structure fabricated at $V_{acc}$ = 1 kV, $I_{beam}$ = 340 pA, and $\tau_{dwell}$ = 1.0 μs, before and after plasma exposure. The decrease in the C Kα peak and the increase in the Pt Mα peak indicate an enhancement of metal purity following treatment.* ***(B)*** *Boxplot of post-treatment $f_{Pt}$ values grouped by $I_{beam}$, showing that dwell time remains a marginal factor in influencing the final composition, as in the pre-treatment case.* ***(C)*** *Post-treatment $f_{Pt}$ values fitted with the same hindered exponential growth model as in the pre-treatment analysis. A substantial increase in $f_{Pt}$ is observed at lower $V_{acc}$, while the change is progressively reduced at higher $V_{acc}$.* ***(D)*** *Relative increase of $f_{Pt,Sat}$ as a function of $V_{acc}$ (left axis, blue), compared with the post-treatment (orange curve) and pre-treatment (grey dashed curve) trends of $f_{Pt,Sat}$ as a function of $V_{acc}$ (right axis).*

For all investigated voltages, the extracted values of the growth constant $k$ are higher post-treatment than their pre-treatment counterparts. However, this increase in $k$ is merely a consequence resulting from the non-uniform enhancement of $f_{Pt}$ across different $I_{beam}$ values.

| $V_{acc}$ [kV] | $f_{Pt,Sat}$ | $k$ [pA$^{-1}$] |
|---|---|---|
| 1 | 0.4953 (0.4552, 0.5354) | 0.0895 (0.0571, 0.122) |
| 3 | 0.3005 (0.2875, 0.3135) | 0.0275 (0.0233, 0.0317) |
| 5 | 0.2309 (0.2081, 0.2537) | 0.0132 (0.0094, 0.0169) |

***Table 2. Hindered exponential growth fitting model after exposure to oxygen plasma.*** *Extracted values of $f_{Pt,Sat}$ and k for each value of $V_{acc}$. In parentheses the 95% confidence bounds are reported.*

**Figure 3D** presents the relative increase of $f_{Pt,Sat}$ as a function of $V_{acc}$ (left axis, blue curve) and compares the post-treatment trend of $f_{Pt,Sat}$ (right axis, orange curve) with its pre-treatment values (right axis, grey dashed curve). Similar to the post-treatment behavior of $k$, the rational function fit in **Equation 2** appears to indicate a steeper decline of $f_{Pt}$ with increasing $V_{acc}$ after treatment. However, this effect is an artifact caused by the non-linear increase of $f_{Pt,Sat}$ across different beam conditions. In contrast, the evolution of $f_{Pt,Sat}$ after the exposure to oxygen plasma reveals a strong dependence on the initial deposition conditions, particularly the acceleration voltage, with the relative increase of $f_{Pt,Sat}$ exhibiting a linear decrease with increasing $V_{acc}$. Despite the initial content of C being lower for low $V_{acc}$, plasma purification is more effective for structures deposited with lower voltages. This behavior can be attributed to differences in C bonding strength rather than absolute C concentration. At low $V_{acc}$, C is incorporated in a loosely bound, amorphous state, making it highly susceptible to oxidation and removal during plasma exposure. In contrast, at high $V_{acc}$, stronger precursor fragmentation leads to the formation of more stable graphitic or carbide-like C structures, which are more resistant to plasma treatment.

### 2.3. Quantification of Morphological Effects

While EDX analysis confirms an increase in the Pt fraction after plasma treatment, it does not provide insight into how this compositional change impacts the structural integrity in EBID-deposited features. Given that oxygen plasma selectively removes C from the deposits, volume shrinkage is expected.

To investigate these effects, a structured test sample is fabricated, consisting of a 5 × 2 matrix of circular openings, each approximately 10 µm in diameter, patterned by FIBM through a 50 nm gold layer with a 5 nm chromium adhesion layer. Each row contains a centrally deposited nanopillar, with the first two rows featuring nanopillars of 100 nm base diameter and the third row containing nanopillars with a 175 nm base diameter. All the nanopillars were fabricated using $V_{acc} = 1$ kV, $I_{beam} = 86$ pA, and $\tau_{dwell} = 1$ µs. **Figure 4A** presents an SEM image of the fabricated nanopillar matrix, while **Figure 4B** provides a magnified view of a 100 nm nominal diameter pillar (highlighted in red in Figure 4A) before and after plasma treatment. After plasma treatment, the nanopillar retains its overall shape but exhibits a noticeable reduction in both height and lateral dimensions. The red dashed lines offer a direct comparison of the height, with the shaded red area representing the height difference $\Delta h$, while lateral shrinkage is also apparent. **Figure 4C** presents grayscale intensity profiles extracted along the green line in Figure 4B. The top plot corresponds to the lateral grayscale profile before oxygen plasma, from which we extracted a diameter of $d_i = 93$ nm, while the bottom plot represents the same profile after oxygen plasma, that showed $d_f = 81$ nm (the pedices '*i*' and '*f*' refer, respectively, to '*initial*' and '*final*'). **Figure 4D** outlines the image processing workflow used to extract the pillar height. First, representative regions are sampled to determine the average grey level of the pillar and the background. A threshold is then applied to generate a binary mask, and the pillar height is extracted from the processed image via a geometrical approximation that considers the volume of a single nanopillar approximated as a cylinder with a hemispherical cap, as schematized in **Figure 5A**.

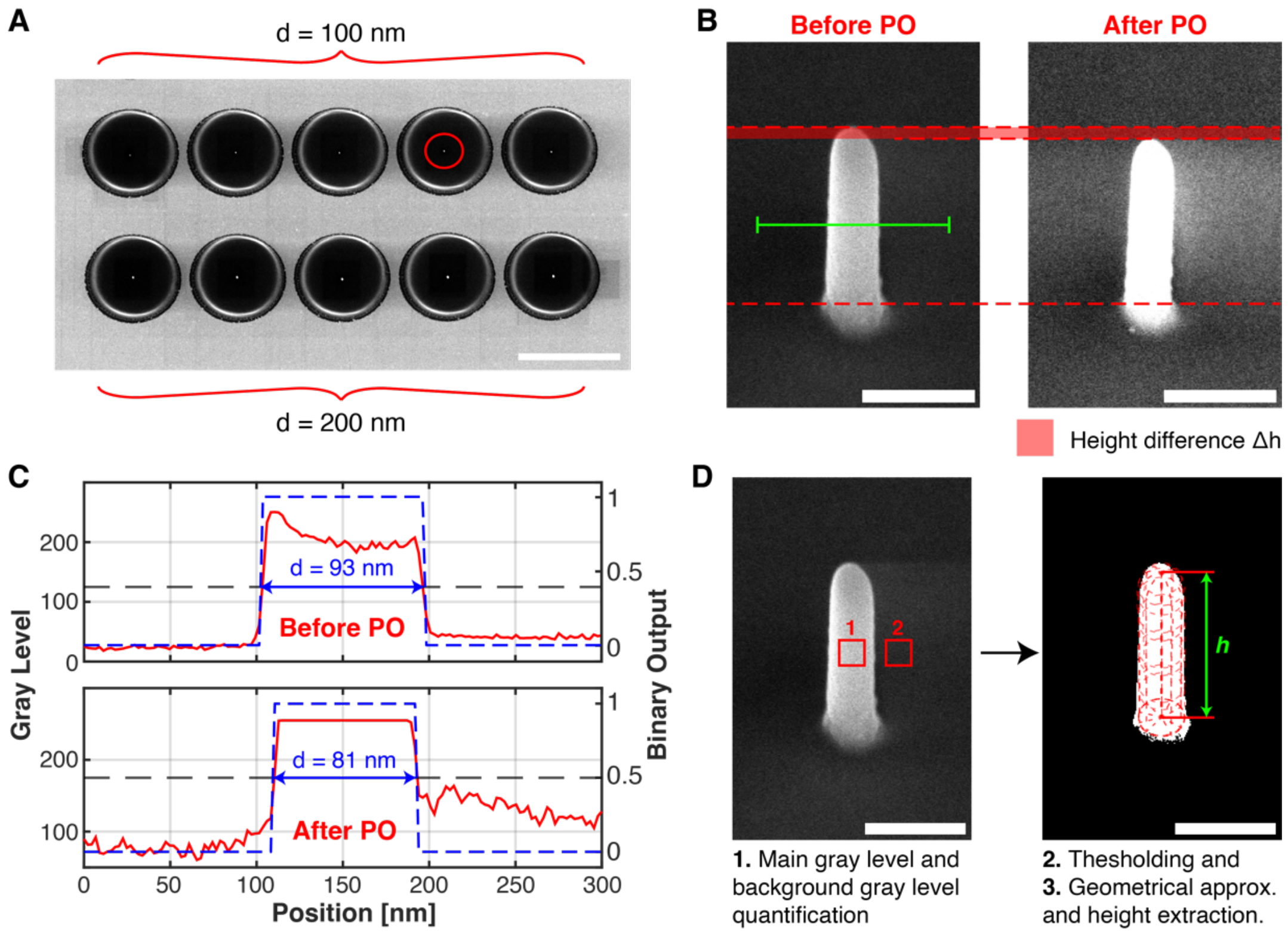

***Figure 4. Morphological changes in EBID-deposited nanopillars following oxygen plasma (PO) treatment. (A)*** *SEM image of the fabricated nanopillar matrix, consisting of a 5 × 2 array of circular openings (~10 µm diameter) milled by FIBM through a 50 nm gold / 5 nm chromium layer, each containing a centrally deposited nanopillar. The red circle highlights the pillar selected for further analysis. The scale bar is 10 µm.* ***(B)*** *Magnified SEM images of the selected nanopillar before and after plasma treatment, showing a reduction in both height and lateral dimensions. The red dashed lines indicate the original and post-treatment heights, with the shaded red area representing the height difference Δh. The scale bars are 200 nm.* ***(C)*** *Grayscale intensity profiles extracted along the green line in panel B, comparing lateral size before (top) and after (bottom) plasma treatment. The extracted diameter (d) is indicated in blue.* ***(D)*** *Image processing workflow for height extraction. Step 1: Mean gray level values are computed for the pillar body and background using user-defined sampling regions. Step 2: A threshold is applied to the grayscale SEM image to generate a binary mask corresponding to the pillar. Step 3: The height (h) is extracted by a simplified geometric model. The scale bar is 200 nm.*

We measure $h_i$ = 364 nm, and $h_f$ = 343 nm, with which the initial and final volumes can be calculated as:

$$\mathcal{V}_{i,f} = \pi r_{i,f}^2 h_{cyl,i,f} + \frac{2}{3}\pi r_{i,f}^3, \tag{5}$$

where $r_{i,f} = d_{i,f}/2$ is the nanopillar radius and $h_{cyl,i,f} = h_{i,f} - r_{i,f}$ is the corrected cylindrical part height subtracting the cap. The initial volume before plasma exposure is therefore $\mathcal{V}_i = 3.02 \times 10^6$ nm$^3$, and the final volume after the treatment is $\mathcal{V}_f = 2.08 \times 10^6$ nm$^3$. The structural effect analysis was extended to all the fabricated test nanostructures, and the results are reported in **Figure 5B** and **5C,** which present histograms comparing the initial and final lateral and vertical dimensions of the nanopillars, respectively. In particular, the bar graph in **Figure 5B** presents a histogram directly comparing the initial diameters (light blue) with the final diameters (dark blue) for each pillar, with the corresponding *Δd* values annotated above each bar. In addition, the percentage diameter reduction is shown in red. All structures exhibit a clear reduction in diameter as a result of the oxygen plasma treatment, which leads to an average shrinkage of 15.67% ± 2.74% (mean value ± standard deviation). When the pillars are grouped according to their initial diameter, the average percentage reduction is 13.57% ± 2.47% for the smaller pillars and 17.77% ± 0.39% for the larger ones. In absolute terms, this corresponds to average diameter losses of 12.18 nm ± 2.29 nm and 30.82 nm ± 0.87 nm for small and large pillars, respectively, suggesting that the effect is more pronounced and uniform in pillars with a larger footprint. **Figure 5C** shows the analogous comparison for vertical dimensions, where initial and final heights (light and dark blue, respectively) are also paired with '$\Delta h$' labels and percentage height reductions (in red). Here, the shrinkage is more moderate and consistent across the population, with an average vertical reduction of 4.91 % ± 1.19 %, and no statistically significant difference between small and large pillars ($t = 1.35$, $p = 0.23$; two-sample t-test).

The observation that lateral shrinkage depends on initial diameter, while vertical shrinkage does not, indicates an anisotropic and geometry-sensitive etching mechanism. This is somewhat counterintuitive, as oxygen plasma is generally expected to interact isotropically with the surface of the sample. A similar anisotropic behavior was observed in gold focused-EBID pillars, where oxygen plasma caused a 20% width reduction but only 10% in height (Supplementary Figure S3 in Ref. [14]), supporting the interpretation that lateral shrinkage may dominate despite the isotropic nature of the effect. While the precise origin of the behavior remains to be clarified, several hypotheses can be proposed. One possibility is that larger pillars, having greater lateral surface area, are more accessible to reactive oxygen species, enhancing the effect's efficiency along the sidewalls. Differences in surface curvature may also play a role: smaller pillars exhibit higher curvature, which may locally modify reaction rates or plasma sheath behavior in ways that reduce lateral erosion. It is also plausible that as C is removed, the increasing Pt concentration near the surface leads to partial passivation, limiting further etching. Finally, geometric factors such as aspect ratio may influence how plasma interacts with different regions of the pillar. The dimensions measured in **Figures 5B** and **5C** were used to calculate the initial and final volumes of all pillars using **Equation 6**.

In the following, we compare the volume change to the change in the Pt fill factor. If we attribute the volume reduction solely removal of C, the Pt fill factor after plasma treatment $f_{Pt,f}$ will result:

$$f_{Pt,f} = f_{Pt,i} \cdot \frac{\mathcal{V}_i}{\mathcal{V}_f}, \qquad (6)$$

where $f_{Pt,i}$ is the initial Pt fill factor, assumed to be $f_{Pt,i} = 0.34$, as determined from the analysis in **Figure 2** for the fabrication parameters used.

**Figure 5D** plots the final volumes as a function of the initial volumes (red dots), and compares them with the values expected under the assumption that volume reduction is solely due to C removal (blue dots), based on an initial Pt fill factor of $f_{Pt,i} = 0.34$ and a final value of $f_{Pt,f} = 0.495$, corresponding to $f_{Pt,Sat}(V_{acc} = 1$ kV). Both data sets were fitted using linear regression, showing excellent agreement with a linear trend. The experimental data indicate a volume reduction described by $V_f = 0.647 \times V_i$ ($R^2 = 0.997$), which is only slightly more pronounced than the expected relationship of $V_f = 0.686 \times V_i$. A dashed reference line representing $V_f = V_i$ is also shown in the figure to highlight the extent of the volume contraction induced by the plasma treatment.

Despite the small absolute discrepancy of ~0.039 between the experimental volume reduction slope and the expected one, the experimental final Pt fill factor $f_{Pt,f} = 0.50 \pm 0.03$ (mean ± standard deviation) remains in excellent agreement with the EDX-derived target value of $f_{Pt,Sat}(1$ kV$) = 0.4953 \pm 0.041$ and fall well within the corresponding 95% confidence interval, as shown in **Figure 5E**. To statistically assess this consistency, we conducted a one-sample t-test under the null hypothesis ($H_0$) that the mean experimental Pt fill factor equals the expected value, against the alternative hypothesis ($H_1$) that the mean differs significantly. The test result ($t = 0.50$, $p = 0.63$) indicates no statistically significant deviation, supporting the conclusion that the observed compositional enrichment is fully compatible with the assumption that the oxygen plasma treatment removes C selectively without significantly affecting the metallic content.

Overall, the combined morphological and compositional analysis demonstrates that the shrinkage induced by plasma treatment is both predictable and compositionally consistent with selective C removal. The excellent agreement between experimentally determined Pt volume fractions and those inferred from geometric measurements supports the validity of using volume-based reconstructions to estimate purification efficiency. This approach provides a powerful, non-destructive tool to monitor compositional changes in FEBID nanostructures when direct compositional analysis (e.g., EDX) is unavailable or impractical, especially in high-aspect-ratio or spatially complex features.

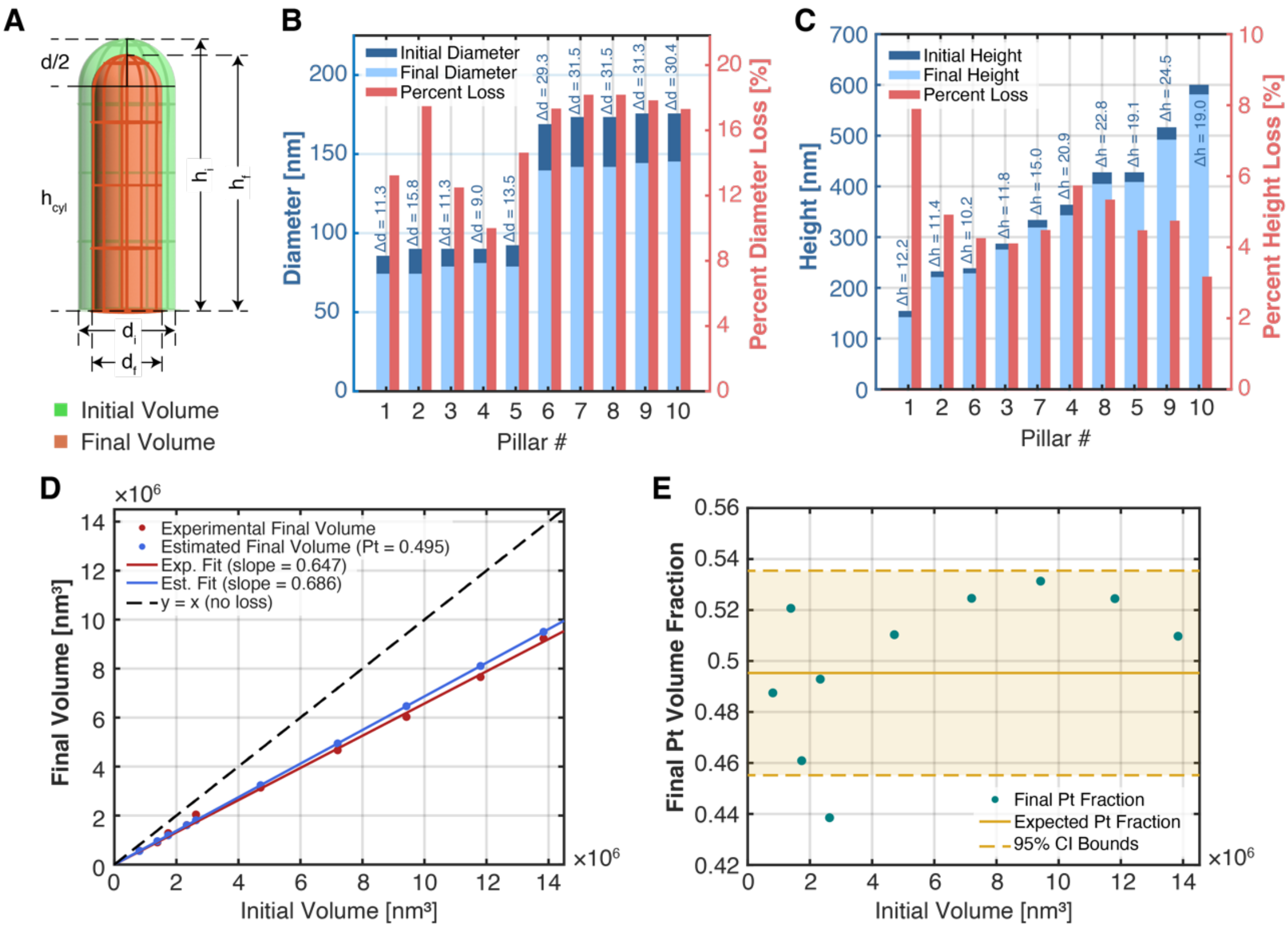

***Figure 5. Quantitative analysis of the geometrical shrinkage and Pt enrichment in BEBID nanopillars after oxygen plasma treatment. A)*** *Schematic representation of a nanopillar with a hemispherical cap, showing the initial and final dimensions used to compute volume.* ***B)*** *Histogram of diameters before and after treatment. Light blue bars represent the final diameter, dark blue the initial diameter, the labels in dark blue the absolute loss Δd, and red bars (right axis) the percentual diameter loss.* ***C)*** *Analogous histogram for pillar heights, showing final height (light blue), initial height (dark blue), and percent height reduction (red, right axis). The labels in dark blue the absolute loss Δh.* ***D)*** *Experimental and estimated final volumes plotted as a function of initial volume. Red points and curve represent the measured data, while blue shows the predicted values assuming Pt conservation and a final Pt volume fraction of 0.495. The dashed line indicates the ideal y = x (no loss).* ***E)*** *Final experimental Pt volume fraction derived from volume conservation, plotted against initial volume. The shaded region marks the 95% confidence interval for the expected Pt fraction (0.4953 ± 0.04).*

## 3. Discussion and Conclusion

The objective of this work is to provide a comprehensive characterization of micro- and nanostructures deposited via Electron Beam Induced Deposition (EBID), using the Pt precursor $MeCpPtMe_3$. In addition, we aim to investigate the effects of post-growth oxygen plasma treatment, implemented using a tabletop system commonly available in many research laboratories and fabrication facilities. This treatment is characterized both in terms of its compositional impact and its morphological effects on representative nanostructures. The characterization of the deposited material, aimed at understanding the influence of fabrication parameters on the metallic fraction, was carried out through Energy Dispersive X-ray Spectroscopy (EDX). Specifically, the effects of beam current $I_{beam}$, accelerating voltage ($V_{acc}$), and dwell time ($\tau_{dwell}$) were investigated. Overall, across all parameter combinations explored, EDX measurements showed that the metallic content increases with beam current, up to a plateau beyond which further increases in current yield no significant enhancement in metal incorporation. At constant current, the metal fraction was found to decrease with increasing accelerating voltage. Conversely, dwell time exhibited a negligible and non-systematic effect on composition, with no clear functional dependence observed. We demonstrated that, at a fixed accelerating voltage ($V_{acc}$), the dependence on beam current follows a hindered exponential growth model. This allows us to estimate both the maximum achievable metal fraction (referred to as the saturation level) and how rapidly it is approached as the current increases. From the overall analysis of all fabricated structures, we also extracted the dependence of the model parameters on the accelerating voltage. This, in principle, enables the formulation of a more comprehensive model that simultaneously accounts for the combined influence of both beam voltage and current on the resulting composition. This characterization was developed here for the Pt precursor $MeCpPtMe_3$, but the same strategy could be extended to other types of precursors, with the goal of expanding the availability of predictive models for the composition of EBID structures across a broader range of metals. We then demonstrated that the metallic fraction can be significantly increased by simply exposing the fabricated sample to a 30-minute oxygen plasma treatment. From a compositional standpoint, we observed that the enhancement depends on the accelerating voltage used during deposition, with the most pronounced improvement occurring at lower $V_{acc}$ values. Specifically, an increase of up to 50% in the metal fraction was measured for $V_{acc} = 1$ kV. Furthermore, we observed that the relative increase in Pt content following purification decreases linearly with increasing accelerating voltage. From a morphological perspective, SEM inspection revealed a volumetric shrinkage fully consistent with the removal of the carbonaceous component, as observed in a set of purposefully fabricated nanostructures. The analysis also indicated that the shrinkage is slightly anisotropic, with a more pronounced effect on the lateral dimensions compared to the axial height, possibly influenced by the aspect ratio of the structures. At this stage, this remains a working hypothesis, and future investigations should aim to systematically explore this behavior across a broader range of aspect ratios and lateral dimensions.

## 4. Materials and Methods

### 4.1. EBID Fabrication of Platinum Nanostructures and Plasma Oxygen Treatment

Nanostructure fabrication was performed using a dual-beam Focused Ion Beam/Scanning Electron Microscope (FIB/SEM), FEI Helios NanoLab 600i, equipped with a gas injection system (GIS). Fabrication commenced once the chamber reached a target vacuum level of $\approx 5 \times 10^{-7}$ mbar.

Firstly, a 170-µm-thick soda lime glass (SLG) coverslip was coated with a ~100 nm gold layer via electron beam evaporation, with a 5 nm chromium adhesion layer. Three 4 × 4 arrays of 5 × 5 $\mu m^2$ apertures on the gold/chromium thin film were obtained using the FIB milling modality, with an accelerating voltage of $V_{acc}^{Ga+} = 30\ kV$ and a beam current of $I_{beam}^{Ga+} = 0.24\ nA$ for a depth of around 300 nm, requiring around 3 minutes per square.

Then, for the EBID process, the GIS needle was introduced in the chamber allowing the Pt precursor MeCpPtMe3 (Trimethyl [(1,2,3,4,5-ETA.)-1 Methyl 2, 4-Cyclopentadien-1-YL] Platinum) to flow. After initiating the gas flow, the chamber pressure was allowed to stabilize at 0.8–1.2 × $10^{-5}$ mbar, maintaining the precursor temperature at ≈49 °C to reduce variations in the growth rate. Electrons were generated through a Schottky field emission gun. For the different fabrication used in the paper, the accelerating voltage was set at 1 kV, 3 kV, 5 kV, and 10 kV, while beam currents of 5.4 pA, 23 pA, 86 pA, and 340 pA were used. Upon completion, to allow the vacuum was paused to return to the level measured before opening the GIS, preventing unintended deposition from residual precursor gas exposure to the SEM beam.

Plasma oxygen treatment was performed using a Harrick Plasma Cleaner (model PDC-002). The device operates with an input voltage of 230 VAC at 50 Hz, and a maximum RF power of 30 W. After the first EDX analysis, the sample was treated in oxygen plasma for 30 minutes at maximum power (30 W) prior to performing the second EDX analysis.

### 4.2. Structural and Compositional Characterization

Imaging and elemental analyses were carried out using a field emission scanning electron microscope (FE-SEM, Ultra55). The microscope is equipped with a Bruker QUANTAX energy-dispersive X-ray spectroscopy (EDX) system, operated with Esprit software, which enables hyperspectral mapping (Hypermap), line profiles, and spot analyses. Elemental quantification was performed using the φ(ρZ) method. The operating conditions for EDX analyses were set to an accelerating voltage of 15 kV and a working distance of 7.5 mm, in order to optimize both spatial resolution and X-ray signal quality.

## 5. Data Availability Statement

The data supporting the findings of this study are currently being curated and will be made available in a public repository upon publication of the article. Reasonable requests for access to the data prior to publication can be directed to the corresponding author.

## 6. Conflict of Interest Disclosure

The authors declare no conflict of interest.

## 7. Acknowledgments

The authors thank Dr. Imène Esteve (Institut de Minéralogie, de Physique des Matériaux et de Cosmochimie, IMPMC, Sorbonne Université/CNRS, Paris) for her assistance with the EDX measurements and analysis. A.Ba. acknowledges funding from the European Union's Horizon 2020 research and innovation program under the Marie Sklodowska-Curie grant agreement (#101106602). A.Ba., M.D.A., A.Br., and A.Ba. also acknowledge funding from the European Union's Horizon 2020 research and innovation program under a grant agreement (#828972). This work was supported by the Plan France 2030 through the project ANR-22-PETQ-0013 and by the French Agence Nationale de la Recherche through the ANR-JCJC2023 program. L.P. is funded by the European Union (ERC, MINING, 101125498). Views and opinions expressed are however those of the author only and do not necessarily reflect those of the European Union or the European Research Council. Neither the European Union nor the granting authority can be held responsible for them. A.Br. is a member of the Institut Universitaire de France (IUF). H.L.J., A.Br., and A. Ba contributed equally to this work and are co-last authors.